\documentclass[preprintnumbers,prd,showpacs,floatfix,superscriptaddress,nofootinbib,twocolumn]{revtex4-1}
\pdfoutput=1
\usepackage{graphicx,epsfig}
\usepackage{amsmath}
\usepackage {amssymb}
\usepackage {longtable}
\usepackage{multirow}
\usepackage{dcolumn}
\usepackage{bm}
\usepackage{amsfonts}
\usepackage{subfigure}
\usepackage{color}
\usepackage{relsize}

\begin{document}

\title{Double gravitational layer traversable wormholes in hybrid metric-Palatini gravity}

\author{Jo\~{a}o Lu\'{i}s Rosa}
\email{joaoluis92@gmail.com}
\affiliation{Institute of Physics, University of Tartu, W. Ostwaldi 1, 50411 Tartu, Estonia}

\date{\today}

\begin{abstract} 
In this work, we explore the existence of traversable wormhole solutions supported by double gravitational layer thin-shells and satisfying the Null Energy Condition (NEC) throughout the whole spacetime, in a quadratic-linear form of the generalized hybrid metric-Palatini gravity. We start by showing that for a particular quadratic-linear form of the action, the junction conditions on the continuity of the Ricci scalar $R$ and the Palatini Ricci scalar $\mathcal R$ of the theory can be discarded without the appearance of undefined distribution terms in the field equations. As a consequence, a double gravitational layer thin-shell arises at the separation hypersurface. We then outline a general method to find traversable wormhole solutions satisfying the NEC at the throat and provide an example. Finally, we use the previously derived junction conditions to match the interior wormhole solution to an exterior vacuum and asymptotic flat solution, thus obtaining a full traversable wormhole solution supported by a double gravitational layer thin-shell and satisfying the NEC. Unlike the wormhole solutions previously obtained in the scalar-tensor representation of this theory, which were scarce and required fine-tuning, the solutions obtained through this method are numerous and exist for a wide variety of metrics and actions. 
\end{abstract}


\maketitle

\section{Introduction}\label{sec:intro}

Wormholes are topological objects connecting two spacetime manifolds. Within general relativity (GR), exact solutions describing these objects were found connecting two differen asymptotic flat spacetimes \cite{morris1,visser1,visser2,visser3} or two different asymptotically de Sitter (dS) or anti-de Sitter (AdS) regions \cite{lemos1}. An essential feature of a wormhole spacetime is a throat satisfying the so-called flaring out-condition. In the context of GR, this feature implies the violation of the Null Energy Condition (NEC), which states that $T_{ab}k^ak^b\geq 0$, where $T_{ab}$ is the matter stress-energy tensor and $k^a$ is an arbitrary null vector. When such a violation happens, matter is denoted as exotic matter. 

In the context of modified theories of gravity, wormhole solutions have also been obtained \cite{agnese1,nandi1,bronnikov1,camera1,lobo1,garattini1,lobo2,garattini2,lobo3,garattini3,sunny}, see also \cite{lobo4} for a review. It has been shown that in this context non-exotic matter can sustain the wormhole throat open, and it is the higher-order curvature terms, which may be interpreted as a gravitational fluid, that support these non-standard wormhole geometries. More precisely, it was shown explicitly that in $f\left(R\right)$ gravity wormhole throats can be constructed without recurring to exotic matter \cite{lobo5}, as well as non-minimal couplings \cite{garcia1,garcia2}, and more generic modified theories of gravity with extra fundamental fields \cite{harko1}. The same kind of solutions were also found in Einstein-Gauss-Bonnet gravity \cite{bhawal1,dotti1,mehdizadeh1}, Brans-Dicke gravity \cite{anchordoqui1}, brane-world scenarios \cite{lobo6}, and the recently proposed hybrid metric-Palatini gravity \cite{capozziello1}.

Indeed, a promising approach to modified gravity consists in having a hybrid metric-Palatini gravitational theory \cite{harko2}, which consists of adding to the Einstein-Hilbert action $R$, a new term $f(\cal{R})$, where $\cal{R}$ is a curvature scalar defined in terms of an independent connection, and $f$ is an arbitrary function of $\cal{R}$. In this approach, the metric and affine connection are regarded as independent degrees of freedom. In this theory, besides wormhole solutions, solar system tests and cosmological solutions have been analyzed  \cite{capozziello2,capozziello3,capozziello4,capozziello5,capozziello6,edery}, see also \cite{harko3} for a review.  The generalized hybrid metric-Platini gravity arises as a natural generalization of the $R+f({\cal R})$ theory, and consists of considering an action $f(R,{\cal R})$ dependent on a general function of both the metric and Palatini curvature scalars \cite{tamanini1}. This class of theories have been shown to provide viable cosmological models \cite{rosa3,rosa4,Rosa:2021ish}, to reproduce the cosmological meta-stability of astrophysical black-holes \cite{rosa5}, the weak-field regime of the theory was also explored \cite{bombacigno,Rosa:2021lhc}, as well as solutions for thick-branes with internal structure \cite{rosa6} and cosmic string-like objects \cite{Rosa:2021zbk}.

In the vast majority of the wormhole literature referenced, although the higher-order curvature terms can be used to force the NEC to be satisfied at the throat of the wormhole, the same condition is violated elsewhere, and thus not the entire spacetime is free of exotic matter. A remarkable exception is the case of the generalized hybrid metric-Palatini gravity where it has been shown that traversable wormhole solutions satisfying the NEC for the whole spacetime can be obtained \cite{rosa1}. These solutions are supported by a thin-shell and matched to an exterior asymptotically-AdS via the use of the theory's junction conditions. However, these solutions carry an important drawback: fine-tuning is required for the solution to satisfy both the NEC and the system of junction conditions, due to the complexity of the latter.

The junction conditions in GR \cite{darmois,lichnerowicz,Israel:1966rt,papa,taub}imply, for a smooth matching between two spacetimes, that the induced metric and the extrinsic curvature must be continuous across the separation hypersurface. These conditions have been used to derive new solutions for the Einstein's field equations, such as constant density stars with an exterior schwarzschild, the Openheimer-Snyder stellar collapse \cite{oppenheimer}, and the matching between FLRW spacetimes with Vaidya, and consequently, Schwarzschild, exteriors \cite{senovilla0}. The matching between two spacetimes can still be done if the extrinsic curvature is discontinuous across the separation hypersurface, but implies the existence of a thin-shell at the junction radius \cite{Israel:1966rt,lanczos1,lanczos2}.  Thermodynamic properties of these shells were introduced \cite{Martinez:1996ni} and the respective entropy was explored in a wide variety of situations, e.g., as rotating shells \cite{Lemos:2017mci,Lemos:2017aol} and electrically charged shells \cite{Lemos:2015ama,Lemos:2016pyc}. Colisions of spacetimes with two shells have also been studied with numerical methods \cite{brito}, as well as stable generalizations of the Schwarzschild stiff fluid star with compactness arbitrarily close to that of a black-hole \cite{rosa7}.

In the context of modified theories of gravity, each theory will present its own set of junction conditions, which must be derived from the respective field equations and the equations of motion of the extra fields, if any. In particular, the junction conditions have been deduced for $f\left(R\right)$ theories of gravity with \cite{Vignolo:2018eco} and without torsion \cite{senovilla1,Deruelle:2007pt}, extensions of the $f\left(R,\right)$ theory e.g. the $f\left(R,T\right)$ gravity \cite{Rosa:2021teg}, scalar-tensor theories \cite{Barrabes:1997kk,suffern}, and also Gauss-Bonnet gravity \cite{Davis:2002gn}.

Similarly to what happens in $f\left(R\right)$ gravity and $f\left(R,T\right)$ gravity, in the generalized hybrid metric-Palatini gravity the system of junction conditions can be simplified for particular forms of the function $f$ \cite{senovilla1,Rosa:2021teg}. As a consequence, new structures called gravitational double-layers appear at the matching hypersurface \cite{senovilla2,eiroa1}. The objective of this paper is to consider a particular form of the function $f\left(R,\mathcal R\right)$ for which the system of junction conditions simplifies and obtain solutions for traversable wormholes supported by double-layer thin-shells that not only satisfy the NEC for the whole spacetime but also drop the requirement for fine-tuning.

This paper is organized as follows: in Sec.\ref{sec:equations} we introduce the quadratic form of the generalized hybrid metric-Palatini gravity theory and derive its junction conditions, in Sec.\ref{sec:worms} we derive solutions for traversable wormholes satisfying the NEC at the throat in this theory, in Sec.\ref{sec:matches} we perform a matching of the wormhole solution with an exterior vacuum solution with the help of a double-layer thin-shell at the separation hypersurface, and finally in Sec.\ref{sec:concl} we draw our conclusions.

\section{The generalized hybrid metric-Palatini gravity}\label{sec:equations}

\subsection{Action and field equations}

The generalized hybrid metric-Palatini gravity is described by an action of the form 
\begin{equation}\label{initialaction}
S=\frac{1}{2\kappa^2}\int_\Omega\sqrt{-g}f\left(R,\cal{R}\right)d^4x+\int_\Omega\sqrt{-g}\;{\cal L}_m d^4x,
\end{equation}
where $\kappa^2\equiv 8\pi G/c^4$, where $G$ is the gravitational constant and $c$ is the speed of light, $\Omega$ is the spacetime manifold on which we define a set of coordinates $x^a$, $g$ is the determinant of the metric $g_{ab}$, $f\left(R,\mathcal R\right)$ is an arbitrary function of the Ricci scalar $R=g^{ab}R_{ab}$, where $R_{ab}$ is the Ricci Tensor, and the Palatini scalar curvature $\mathcal R=g^{ab}\mathcal R_{ab}$, where $\mathcal R_{ab}$ is the Palatini Ricci scalar written in terms of an independent connection $\hat\Gamma^c_{ab}$ as 
\begin{equation}\label{act1}
\mathcal{R}_{ab}=\partial_c\hat\Gamma^c_{ab}-\partial_b\hat\Gamma^c_{ac}+\hat\Gamma^c_{cd} \hat\Gamma^d_{ab}-\hat\Gamma^c_{ad}\hat\Gamma^d_{cb},
\end{equation}
and $\mathcal L_m$ is the matter Lagrangian considered minimally coupled to the metric $g_{ab}$. In the following, we shall consider a geometrized system of units in such a way that $G=c=1$ and the constant $\kappa^2$ reduces to $\kappa^2=8\pi$. Equation \eqref{act1} depends on two independent variables, namely the metric $g_{ab}$ and the independent connection $\hat\Gamma^c_{ab}$, and thus one can derive two equations of motion.

Varying Eq.\eqref{act1} with respect to the metric $g_{ab}$ yields the modified field equations
\begin{eqnarray}
\frac{\partial f}{\partial R}R_{ab}+\frac{\partial f}{\partial \mathcal{R}}\mathcal{R}_{ab}-\frac{1}{2}g_{ab}f\left(R,\cal{R}\right)
   \nonumber \\
-\left(\nabla_a\nabla_b-g_{ab}\Box\right)\frac{\partial f}{\partial R}=8\pi T_{ab},\label{eqfield}
\end{eqnarray}
where $\nabla_a$ is the covariant derivative and $\Box=\nabla^a\nabla_a$ the d'Alembert operator,
both written in terms of the Levi-Civita connection $\Gamma^c_{ab}$ of the metric $g_{ab}$, and
$T_{ab}$ is the stress-energy tensor defined in the usual manner as
\begin{equation} \label{stress}
T_{ab}=-\frac{2}{\sqrt{-g}}\frac{\delta(\sqrt{-g}\,{\cal L}_m)}{\delta(g^{ab})}.
\end{equation}

Let us now analyze the variation of Eq.\eqref{act1} with respect to the independent connection $\hat\Gamma^c_{ab}$, which provides the relation
\begin{equation}\label{eqindcon}
\hat\nabla_c\left(\sqrt{-g}\frac{\partial f}{\partial \cal{R}}g^{ab}\right)=0,
\end{equation}
where $\hat\nabla_a$ is the covariant derivative written in terms of the independent connection $\hat\Gamma^c_{ab}$. Noting that $\sqrt{-g}$ is a scalar density of weight 1, and thus $\hat\nabla_a\sqrt{-g}=0$, Eq.\eqref{eqindcon} implies the existence of another metric tensor $h_{ab}=\left(\partial f/\partial\mathcal R\right)g_{ab}$ conformally related to the metric $g_{ab}$ with conformal factor $\partial f/\partial \mathcal R$, for which the independent connection $\hat\Gamma^c_{ab}$ is the Levi-Civita connection, i.e., 
\begin{equation}
\hat\Gamma^a_{bc}=\frac{1}{2}h^{ad}\left(\partial_b h_{dc}+\partial_c h_{bd}-\partial_d h_{bc}\right),
\end{equation}
where $\partial_a$ denotes partial derivatives. This conformal relation between $g_{ab}$ and $h_{ab}$ implies that the two Ricci tensors $R_{ab}$ and $\mathcal R_{ab}$, assumed \textit{a priori} as independent, are actually related to each other via the expression
\begin{equation}\label{ricrel}
\mathcal R_{ab}=R_{ab}-\frac{1}{f_\mathcal R}\left(\nabla_a\nabla_b+\frac{1}{2}g_{ab}\Box\right)f_\mathcal R+\frac{3}{2f_\mathcal R^2}\partial_a f_\mathcal R\partial_b f_\mathcal R,
\end{equation} 
where the subscripts $R$ and $\mathcal R$ denote derivatives with respect to the same variables. Eqs.\eqref{eqindcon} and \eqref{ricrel} are equivalent, and thus we shall consider only the latter for simplicity. 

\subsection{Junction conditions}\label{sec:junct}

\subsubsection{Notation and assumptions}

Let $\Sigma$ be a hypersurface that separates the spacetime $\mathcal
V$ into two regions, $\mathcal V^+$ and $\mathcal V^-$. Let us
consider that the metric $g_{ab}^+$, expressed in coordinates $x^a_+$,
is the metric in region $\mathcal V^+$ and the metric $g_{ab}^-$,
expressed in coordinates $x^a_-$, is the metric in region $\mathcal
V^-$, where the latin indeces run from $0$ to $3$. Let us assume that
a set of coordinates $y^\alpha$ can be defined in both sides of
$\Sigma$, where greek indeces run from $0$ to $2$. The projection
vectors from the 4-dimensional regions $\mathcal V^\pm$ to the
3-dimensional hypersurface $\Sigma$ are $e^a_\alpha=\partial
x^a/\partial y^\alpha$. We define $n^a$ to be the unit normal vector
on $\Sigma$ pointing in the direction from $\mathcal V^-$ to $\mathcal
V^+$. Let $l$ denote the proper distance or time along the geodesics
perpendicular to $\Sigma$ and choose $l$ to be zero at $\Sigma$,
negative in the region $\mathcal V^-$, and positive in the region
$\mathcal V^+$. The displacement from $\Sigma$ along the geodesics
parametrized by $l$ is $dx^a=n^adl$, and $n_a=\epsilon \partial_a l$,
where $\epsilon$ is either $1$ or $-1$ when $n^a$ is a spacelike or
timelike vector, respectively, i.e., $n^an_a=\epsilon$.

We will be working using distribution functions. For any quantity
$X$, we define $X=X^+\Theta\left(l\right)+X^-\Theta\left(-l\right)$,
where the indeces $\pm$ indicate that the quantity $X^\pm$ is the
value of the quantity $X$ in the region $\mathcal V^\pm$, and
$\Theta\left(l\right)$ is the Heaviside distribution function, with
$\delta\left(l\right)=\partial_l\Theta\left(l\right)$ the Dirac
distribution function. We also denote
$\left[X\right]=X^+|_\Sigma-X^-|_\Sigma$ as the jump of $X$ across
$\Sigma$, which implies by definition that
$\left[n^a\right]=\left[e^a_\alpha\right]=0$.

\subsubsection{Constraints on the form of $f\left(R,\mathcal R\right)$}

To match the interior spacetime $\mathcal V^-$ to the exterior spacetime $\mathcal V^+$, we need to use the junction conditions of the generalized hybrid metric-Palatini gravity. These conditions were derived for a general form of the function $f\left(R,\mathcal R\right)$ and already used to obtain wormhole solutions in Ref.\cite{rosa1}. Defining the induced metric at $\Sigma$ as $h_{\alpha\beta}=e_\alpha^ae_\beta^bg_{ab}$ and the extrinsic curvature $K_{\alpha\beta}=e_\alpha^ae_\beta^b\nabla_a n_b$, the junction conditions for a matching between the two spacetimes with a thin-shell of perfect fluid at the separation hypersurface $\Sigma$ can be generically written in the form
\begin{eqnarray}
&\left[h_{\alpha\beta}\right]=0,\label{junction1}\\
&\left[K\right]=0,\label{junction2}\\
&\left[R\right]=0,\label{junction3}\\
&\left[\mathcal R\right]=0,\label{junction4}\\
&f_{\mathcal R R}n^a\left[\partial_aR\right]+f_{\mathcal R\mathcal R}n^a\left[\partial_a\mathcal R\right]=0,\label{junction5}\\
&\epsilon\delta_\alpha^\beta n^c\left[\partial_cR\right]\left(f_{RR}-\frac{f_{R\mathcal R}^2}{f_{\mathcal R\mathcal R}}\right)-\nonumber\\
&-\left(f_R+f_\mathcal R\right)\epsilon\left[K_\alpha^\beta\right]=8\pi S_\alpha^\beta,\label{junction6}
\end{eqnarray}
where $K=K_\alpha^\alpha$ is the trace of the extrinsic curvature and $S_\alpha^\beta$ is the stress-energy tensor of the thin-shell arising at the separation hypersurface $\Sigma$. The conditions in Eqs.\eqref{junction3} and \eqref{junction4} are imposed to avoid the presence of products $\Theta\left(l\right)\delta\left(l\right)$ in the field equations in Eq.\eqref{eqfield}, which are undefined in the distribution formalism. In the following, we pursue an alternative way of avoiding these problematic products by constraining the function $f\left(R,\mathcal R\right)$ instead. 

Let us start by defining properly the metric $g_{ab}$ throughout the spacetime using the distribution formalism. The metric $g_{ab}$ takes the form
\begin{equation}
g_{ab}=g_{ab+}\Theta(l)+g_{ab-}\Theta(-l).
\end{equation}
The Ricci tensor $R_{ab}$ and the Ricci scalar $R$ associated with this metric in the distribution formalism, upon imposition of Eqs.\eqref{junction1} and \eqref{junction2}, become
\begin{equation}\label{district}
R_{ab}=R_{ab}^+\Theta\left(l\right)+R_{ab}^-\Theta\left(-l\right)-\epsilon\left[K_{ab}\right]\delta\left(l\right),
\end{equation}
\begin{equation}\label{distric}
R=R^+\Theta\left(l\right)+R^-\Theta\left(-l\right),
\end{equation}
respectively, where we have used Eq.\eqref{junction2}. Taking the first-order partial derivatives of the Ricci scalar $\partial_a R$ leads to $\partial_a R=\partial_aR^+\Theta\left(l\right)+\partial_aR^-\Theta\left(-l\right)+\epsilon\left[R\right] n_a\delta\left(l\right)$, and a similar expression for the partial derivatives of $\mathcal R$. In the definition of the Palatini Ricci scalar in Eq.\eqref{ricrel} it is evident that, due to the existence of terms $\partial_af_\mathcal R\partial_bf_\mathcal R$, there will be terms depending on products of the form $\partial_aR\partial_bR$ and $\partial_a\mathcal R\partial_b\mathcal R$. These products will lead to terms proportional to $\Theta\left(l\right)\delta\left(l\right)$, which are undefined in the distribution formalism, or $\delta\left(l\right)^2$, which are singular. Thus, for the Palatini Ricci tensor to be properly defined in the distribution formalism, it is thus necessary to eliminate these terms from Eq.\eqref{ricrel}. For a general form of the function $f\left(R,\mathcal R\right)$, the presence of these terms is avoided by imposing Eqs.\eqref{junction3} and \eqref{junction4} as junction conditions, thus canceling all proportionality in $\delta\left(l\right)$ and solving the problem. However, an alternative solution for this problem is to impose that $f_\mathcal R$ is a constant, i.e., the function $f\left(R,\mathcal R\right)$ is linear in $\mathcal R$. Consequently, $\partial_a f_\mathcal{R}=0$ and these products are effectively removed from Eq.\eqref{ricrel}. The function $f\left(R,\mathcal R\right)$ thus becomes
\begin{equation}
f\left(R,\mathcal R\right)=g\left(R\right)+\gamma \mathcal R,
\end{equation}
where $g\left(R\right)$ is a well-behaved function of the Ricci scalar $R$ and $\gamma$ is a constant. On the other hand, due to the existence of differential terms of the form $\nabla_a\nabla_b f_R$ and $\Box f_R$ in the field equations in Eq.\eqref{eqfield}, the same products of the form $\partial_aR\partial_bR$ will also arise from these terms, thus leading to the same problems related to undefined $\Theta\left(l\right)\delta\left(l\right)$ or singular $\delta\left(l\right)^2$ terms. Again, one can avoid the presence of these terms without imposing Eqs.\eqref{junction3} and \eqref{junction4} by considering instead that $f_{RR}$ is constant, i.e., the function $f\left(R,\mathcal R\right)$ is quadratic in $R$. Consequently, $\nabla_a\nabla_b f_R = f_{RR}\nabla_a\nabla_b R$, and thus the problematic products are effectively removed from the field equations. The function $f\left(R,\mathcal R\right)$ can be written in the form
\begin{equation}\label{quadf}
f\left(R,\mathcal R\right)=R-2\Lambda+\alpha R^2+\gamma\mathcal R,
\end{equation}
where the parameters $\Lambda$ and $\alpha$ are constants. Inserting this form of the function $f\left(R,\mathcal R\right)$ into Eq.\eqref{ricrel}, all the differential terms in $f_\mathcal R$ cancel, we obtain $R_{ab}=\mathcal R_{ab}$, and consequently we have $R=\mathcal R$. Furthermore, we have $f_R=1+2\alpha R$, which implies that $\partial_af_R=2\alpha\partial_aR$ and $\nabla_a\nabla_bf_R=2\alpha\nabla_a\nabla_bR$. Under these considerations, the modified field equations given in Eq.\eqref{eqfield} simplify to
\begin{eqnarray}
&&R_{ab}\left(1+\gamma+2\alpha R\right)-2\alpha\left(\nabla_a\nabla_b-g_{ab}\Box\right)R-\nonumber \\
&&-\frac{1}{2}g_{ab}\left[R\left(1+\gamma+\alpha R\right)-2\Lambda\right]=8\pi T_{ab}.\label{quadfield}
\end{eqnarray}
Notice also that for the particular form of $f\left(R,\mathcal R\right)$ in Eq.\eqref{quadf}, since $f_\mathcal R=\gamma$ is a constant, then the partial derivatives $f_{\mathcal R\mathcal R}$ and $f_{\mathcal R R}$ also vanish and the junction condition in Eq.\eqref{junction5} is automatically satisfied.

Let us point a few considerations about the parameters $\alpha$, $\gamma$ and $\Lambda$. As we stand in a particular case where $R=\mathcal R$, the coefficient of $R$ in the function $f$ is effectively $1+\gamma$. As we want to preserve the positivity of the Einstein-Hilbert term in the action, this imposes that $\gamma>-1$. Also, although there are no constraints on the sign of the quadratic term, it has been shown that a positive $\alpha$ provides fruitful results in cosmology, more specifically in models for inflation\cite{starobinski}. One such case is the Starobinski model $\alpha=1/m^2$, where $m$ is a constant with units of mass. We will thus focus on models with $\alpha>0$. Finally, as it is defined, $\Lambda$ plays the role of a cosmological constant and it controls the asymptotics of the solution, which will be de Sitter (if $\Lambda>0$), anti-de Sitter (if $\Lambda<0$), or Minkowski (if $\Lambda=0$).

Since for this particular form of the function $f\left(R,\mathcal R\right)$ in Eq.\eqref{quadf} the Ricci scalar $R$ and the Palatini Ricci scalar $\mathcal R$ coincide, because the conformal factor $f_\mathcal R=\gamma$ is a constant, one should clarify the role of the Palatini term in the analysis and emphasize what are the differences between this situation and a purely $f\left(R\right)$ model. Notice that if one replaces $\mathcal R$ by $R$ in Eq.\eqref{quadf} to obtain a function $f\left(R,\mathcal R\right)$ that depends solely in $R$, the conformal factor $f_\mathcal R$ vanishes and the two metrics $g_{ab}$ and $h_{ab}$ coincide. In our situation, even though $R$ and $\mathcal R$ are forced to be the same by Eq.\eqref{ricrel}, the explicit dependence of $f\left(R,\mathcal R\right)$ in $\mathcal R$ guarantees that the conformal factor $f_\mathcal R=\gamma$ is non-zero and the metrics $g_{ab}$ and $h_{ab}$ differ. Furthermore, one could argue that a redefinition of the parameters of the form $\bar\alpha=\alpha/\left(1+\gamma\right)$ and $\bar\Lambda=\Lambda/\left(1+\gamma\right)$ and a factorization of $\left(1+\gamma\right)$ from the action would lead to an $f\left(R,\mathcal R\right)=f\left(R\right)$ function that produces the same field equations as Eq.\eqref{quadf}. While this is true if the matter Lagrangian $\mathcal L_m$ vanishes, the same does not hold if $T_{ab}\neq 0$, as it would also require a redefinition of the stress-energy tensor as $\bar T_{ab}=T_{ab}/\left(1+\gamma\right)$, and the distribution of matter would only be the same if $\gamma=0$. Thus, the Palatini term in Eq.\eqref{quadf} effectively controls the influence of the matter distribution in gravity: as one increases the value of $\gamma$, less matter is needed to produce the same gravitational effects.

Finally, we note that the generalized hybrid metric-Palatini gravity can often be written in terms of a dynamically equivalent scalar-tensor representation, which can be deduced through the definition of auxiliary fields in the action in Eq.\eqref{act1}. However, the particular form of the function $f\left(R,\mathcal R\right)$ deduced in Eq.\eqref{quadf} does not have an equivalent scalar-tensor representation, see Appendix \ref{sec:app} for more details. Thus, it is sufficient to pursue this analysis in the geometrical representation of the theory.

\subsubsection{Double-layer thin-shells}

In the previous section, we have deduced a particular form of the function $f\left(R,\mathcal R\right)$ that simplifies the set of junction conditions. This quadratic form of the function $f\left(R,\mathcal R\right)$ is provided in Eq.\eqref{quadf}. Let us now analyze the consequences of considering this form of the function.

The Ricci tensor and Ricci scalar have the forms provided in Eqs.\eqref{district} and \eqref{distric}, respectively. Let us now analyse the second-order derivative terms of the Ricci scalar $R$ in the field equations in Eq.\eqref{quadfield}, i.e., the terms $\nabla_a\nabla_bR$ and $\Box R$. To do so, let us take a double covariant derivative of $R$ which becomes
\begin{equation}\label{specialdric}
\nabla_a\nabla_bR=\left(\nabla^2R\right)_{ab}+\epsilon\nabla_a\left(\left[R\right]\delta\left(l\right)n_b\right),
\end{equation}
\begin{equation}\label{derivric2}
\left(\nabla^2R\right)_{ab}=\nabla_a\nabla_bR_+\Theta\left(l\right)+\nabla_a\nabla_bR_-\Theta\left(-l\right)+\epsilon\delta\left(l\right)n_a\left[\partial_bR\right].
\end{equation}
These results imply that the junction conditions $\left[R\right]=0$ and $\left[\mathcal R\right]=0$ can be discarded in the particular case of Eq.\eqref{quadf} but, as a consequence, extra terms will arise in the stress-energy tensor $S_{ab}$ of the thin-shell.

The second term in the right-hand side of Eq.~\eqref{specialdric} is also familiar from the $f\left(R\right)$ gravity case. Following \cite{senovilla1}, we verify that this term can be rewritten in the form
\begin{equation}
\nabla_a\left(\left[R\right]\delta\left(l\right)n_b\right)=\delta\left(l\right)\left(K_{ab}-\epsilon Kn_an_b+n_bh^c_a\nabla_c\right)\left[R\right]+\Delta_{ab},
\end{equation}
where $\Delta_{ab}$ is a new distribution function defined by
\begin{equation}
\int_\Omega \Delta_{ab}Y^{ab}d^4x=-\epsilon\int_\Sigma \left[R\right]n_an_bn^c\nabla_cY^{ab}d^3x,
\end{equation}
for some test function $Y^{ab}$. Inserting this result
into Eq.~\eqref{specialdric} and using Eq.~\eqref{derivric2} we obtain
\begin{eqnarray}
\label{singd2ric}
\nabla_a\nabla_bR=&&\nabla_a\nabla_bR^+
\Theta\left(l\right)+\nabla_a\nabla_bR^-
\Theta\left(-l\right)+\epsilon\Delta_{ab}+
\nonumber
\\
&&+
\epsilon\delta\left(l\right)\left[ \epsilon n_an_bn^c
\left[\nabla_cR\right]+2n_{(a}h^c_{b)}\nabla_c
\left[R\right]+ \right.
\nonumber
\\
&&+
\Big.\left[R\right]\left(K_{ab}-\epsilon K n_an_b
\right)\Big]
\,.
\end{eqnarray}
Consequently, contracting with $g^{ab}$, we obtain
\begin{eqnarray}
&&\label{singboxric}\Box R=\Box R^+\Theta\left(l\right)+\Box R^-\Theta\left(-l\right)+\\ 
&&+\epsilon\Delta+\epsilon\delta\left(l\right)n^c\left[\nabla_cR\right],\nonumber
\end{eqnarray}
where we defined $\Delta=\Delta^a_a$ and we have used $n^an_a=\epsilon$, $\epsilon^2=1$, and $n^ah^c_a=0$. Inserting the results in Eqs.~\eqref{singd2ric} and~\eqref{singboxric} into the field equations given in Eq.~\eqref{quadfield}, considering the distribution of $R_{ab}$ given by Eq.~\eqref{district}, and keeping only the singular terms, we verify that the stress-energy tensor becomes in this case
\begin{eqnarray}\label{tabquad}
T_{ab}&=&T_{ab}^+\Theta\left(l\right)+T_{ab}^-\Theta\left(-l\right)+\\
&+&\delta\left(l\right)\left(S_{ab}+S_{(a}n_{b)}+Sn_an_b\right)+s_{ab}\left(l\right),\nonumber
\end{eqnarray}
where $S_{ab}$ is the usual stress-energy tensor of the thin-shell, $S_a$ is the external flux momentum whose normal component measures the normal energy flux across $\Sigma$ and the spacial components measure the tangential stresses, $S$ measures the external normal pressure or tension supported on $\Sigma$, and $s_{ab}$ is the double-layer stress-energy tensor distribution. These variables are given in terms of the geometrical quantities as
\begin{eqnarray}
8\pi S_{ab}&=&-\epsilon\left[K_{ab}\right]\left(1+\gamma\right)+\label{stressts}\\
&+&2\epsilon\alpha\left(h_{ab}n^c\left[\nabla_cR\right]-R_\Sigma\left[K_{ab}\right]-K_{ab}^\Sigma\left[R\right]\right),\nonumber
\end{eqnarray}
\begin{equation}\label{sa}
8\pi S_a=-2\epsilon \alpha h^c_a\nabla_c\left[R\right],
\end{equation}
\begin{equation}\label{s}
8\pi S=2\epsilon\alpha K\left[R\right],
\end{equation}
\begin{equation}\label{sab}
8\pi s_{ab}\left(l\right) = 2\epsilon\alpha \, \Omega_{ab}\left(l\right),
\end{equation}
Here, $R_\Sigma$ and $K_{ab}^\Sigma$ are the Ricci scalar and the extrinsic curvature of the hypersurface where the thin-shell lives. As these two variables are, in general, discontinuous, one defines $R_\Sigma=\left(R^++R^-\right)/2$ and $K^\Sigma_{ab}=\left(K^+_{ab}+K^-_{ab}\right)/2$, following Ref. \cite{senovilla2}. 
In Eq.~(\ref{sab}), $\Omega_{ab}(l) \equiv g_{ab}^\Sigma\Delta-\Delta_{ab}$ and the
double-layer stress-energy distribution becomes
\begin{equation}\label{sabfinal}
\int_\Omega 8\pi s_{ab} Y^{ab} d^4x=-\int_\Sigma 2\epsilon \alpha 
\left[R\right] h_{ab} n^c\nabla_cY^{ab}d^3x.
\end{equation}

As expected, every extra term that arises in this formalism has a dependence on $\left[R\right]$, and the usual results for the general case can be recovered by setting $\left[R\right]=0$. These gravitational double layers have been studied in the scope of $f\left(R\right)$ gravity for the particular case of timelike hypersurfaces \cite{senovilla2} and our results can be matched to the ones obtained in $f\left(R\right)$ by setting $\gamma=1$ and $\epsilon=1$.

As for the double-layer stress-energy tensor distribution $s_{ab}$ in Eq.~(\ref{sabfinal}), 
it has been noted to have resemblances to classical dipole distrubitions, 
with a dipole strength $8\pi \mathcal P_{ab} = 2\alpha [R] h_{ab}$~\cite{eiroa1},
here arising mainly due to the existence of a nonzero jump in the curvature.
In GR, thin shell spacetimes with $[R] \neq 0$ are also acceptable, but Eqs.~(\ref{sa}-\ref{sab}) do not manifest, since $\alpha = 0$ for GR.
The dipole distribution term in Eq.~(\ref{sab}) is a novelty and so far unique to the quadratic term of $f(R)$ extensions of gravity. However, there is still no physical interpretation for why a dipole distribution should arise in gravity. 
In spite of this, Eq.~(\ref{sab}) is still a crucial piece in assuring that the stress-energy tensor
distribution is divergence free, i.e. $\nabla^a T_{ab}=0$ ~\cite{senovilla1}.

\section{Wormhole ansatz and solutions}\label{sec:worms}

\subsection{General considerations on wormhole spacetimes}

The general metric that describes a static and spherically symmetric wormhole spacetime in the usual spherical coordinates $\left(t,r,\theta,\phi\right)$ is given by the line element
\begin{equation}\label{metric}
ds^2=-e^{\zeta\left(r\right)}dt^2+\left[1-\frac{b\left(r\right)}{r}\right]^{-1}dr^2+r^2d\Omega^2,
\end{equation}
where $\zeta\left(r\right)$ is the redshift function, $b\left(r\right)$ is the shape function, and $d\Omega^2=d\theta^2+\sin^2\theta d\phi^2$ is the solid angle surface element. These two metric functions are not arbitrary. In this work, we are interested in traversable wormhole solutions. For a wormhole spacetime to be traversable, its metric functions must fulfill two conditions. First, the wormhole spacetime must not present any horizons, as so to allow a traveler to move arbitrarily close to the wormhole throat at $r=r_0$ without being prevented to escape the interior region. To fulfill this requirement, the redshift function should be finite throughout the whole spacetime, i.e., $|\zeta\left(r\right)|<\infty$. The second condition is a fundamental geometric condition in wormhole physics at the wormhole throat that guarantees its traversability. The so-called flaring-out condition is translated into two boundary conditions for the shape function given by 
\begin{equation}\label{flareout}
b\left(r_0\right)=r_0,\qquad b'\left(r_0\right)<1.
\end{equation}
Two broad families of redshift and shape functions that satisfy the requirements for wormhole traversability are 
\begin{equation}\label{redshift}
\zeta\left(r\right)=\zeta_0\left(\frac{r_0}{r}\right)^p,
\end{equation}
\begin{equation}\label{shape}
b\left(r\right)=r_0\left(\frac{r_0}{r}\right)^q,
\end{equation}
where $\zeta_0$ is a dimensionless constant, $r_0$ is the radius of the wormhole throat, and $p$ and $q$ are constant exponents.

Let us now turn to the matter sector. We assume that matter is described by an anisotropic perfect fluid, i.e., the stress-energy tensor can be written as
\begin{equation}\label{tabani}
T_a^b=\text{diag}\left(-\rho,p_r,p_t,p_t\right),
\end{equation}
where $\rho=\rho\left(r\right)$ is the energy density, $p_r=p_r\left(r\right)$ is the radial pressure, and $p_t=p_t\left(r\right)$ is the transverse pressure, all assumed to depend only in the radial coordinate as to preserve the spherical symmetry and staticity of the spacetime. Furthermore, we are interested in matter that satisfies the NEC. For the NEC to be satisfied, $\rho$, $p_r$ and $p_t$ must satisfy the following inequalities:
\begin{equation}\label{nec}
\rho+p_r>0\qquad \rho+p_t>0.
\end{equation}
Within general relativity, the flaring out condition, Eq.\eqref{flareout}, and the NEC, Eq.\eqref{nec}, are incompatible. Indeed, Eq.\eqref{flareout} is effectively a condition on the Einstein's tensor $G_{ab}$, which is directly related to the stress-energy tensor $T_{ab}$ via the Einstein's field equations $G_{ab}=8\pi T_{ab}$, resulting in a condition for the latter as $T_{ab}k^ak^b<0$, for any null vector $k^a$, which is equivalent to a violation of Eq.\eqref{nec} in the particular case of an anisotropic perfect fluid. In modified gravity, the modified field equations can be rewritten in the form $G_{ab}=8\pi T_{ab}^{\text{eff}}$, where $T_{ab}^{\text{eff}}$ is the effective stress energy tensor that includes not only the matter $T_{ab}$ but also the higher-order curvature terms from Eq.\eqref{quadfield}. In this context, it is the effective stress-energy tensor that must fulfill the condition $T_{ab}^{\text{eff}}k^ak^b<0$, and thus it is possible, in principle, that Eq.\eqref{nec} is verified as long as the extra higher-order curvature terms compensate the positive matter contributions.

\subsection{wormhole solutions satisfying the NEC at the throat}\label{sec:wormsol}

The general strategy to obtain wormhole solutions is as follows. We take the field equations given in Eq.\eqref{quadfield}, we select a metric of the form provided in Eq.\eqref{metric} with redshift and shape functions defined as in Eqs.\eqref{redshift} and \eqref{shape}, respectively, and we use the stress-energy tensor of an anisotropic perfect fluid as shown in Eq.\eqref{tabani}. Given the spherical symmetry of the problem, there will be three independent field equations, which we do not write explicitly due to their lenghty character. These three equations are then solved to the three independent unknowns, $\rho$, $p_r$ and $p_t$. Having obtained the solutions for the matter fields, one computes the left-hand side of the NEC in Eq.\eqref{nec} and verifies if this condition is satisfied at the throat $r=r_0$. Different wormhole models can be tested by varying the exponents $p$ and $q$ and constants $r_0$ and $\zeta_0$, as well as different actions by varying the parameters $\alpha$, $\gamma$ and $\Lambda$. 

\begin{figure}[h!]
\includegraphics[scale=0.8]{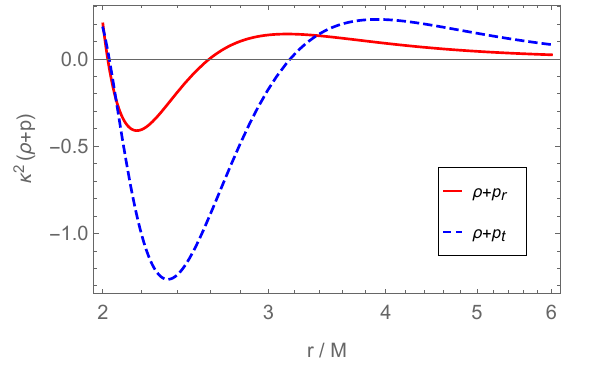}
\caption{Plot of $\rho+p_r$ (solid red line) and $\rho+p_t$ (dashed blue line) for $\alpha=0.15/M^2$, $\gamma=10$, $\Lambda=0$, $\zeta_0=-60$, and $r_0=2M$, where $M$ is a constant with units of mass. The NEC given in Eq.\eqref{nec} is satisfied at the throat $r=r_0$ but violated elsewhere.}
\label{fig:nec}
\end{figure}

As an example, let us consider a model with redshift and shape functions inversely proportional to the radius $r$, i.e. $p=q=1$, let us consider that $\Lambda=0$ to preserve asymptotic flatness, and let us consider the throat to be at the Schwarzschild radius $r_0=2M$.  Following the procedure outlined in the previous paragraph, one verifies that there are many different combinations of the parameters $\alpha$, $\gamma$, and $\zeta_0$ for which the NEC is satisfied at the throat. In Fig.\ref{fig:nec}, we plot one such solution. one verifies that the NEC given in Eq.\eqref{nec} is satisfied at the throat of the wormhole $r=r_0$, but it is eventually violated at larger values of $r$. For this solution to satisfy the NEC throughout the whole spacetime, one needs to perform a matching with an exterior vaccum solution in the region where the NEC is still satisfied. We now turn to this.

\section{Matching with an exterior vacuum solution}\label{sec:matches}

\subsection{Selecting an exterior solution}

In this section, we want to select an exterior solution to be matched with an interior wormhole solution derived within the framework outlined in Sec.\ref{sec:wormsol}. As we want the exterior solution to satisfy the NEC, the simplest way of doing so is to consider a vacuum exterior solution, i.e., with $T_{ab}=0$. For simplicity, we also want to preserve asymptotic flatness, and thus we impose $\Lambda=0$. Inserting these assumptions into Eq.\eqref{quadfield}, one verifies that any solution in GR with $R_{ab}=0$ (and consequently $R=0$) will also be a solution of the modified field equations. As the solutions computed for wormhole interiors in Sec.\ref{sec:wormsol} are spherically symmetric, we shall take the exterior vacuum solution to be a Schwarzschild spacetime, i.e., the line element in standard spherical coordinates is
\begin{equation}\label{schwarz}
ds^2=-\left(1-\frac{2M}{r}\right)e^{\zeta_e}dt^2+\left(1-\frac{2M}{r}\right)^{-1}dr^2 + r^2d\Omega^2,
\end{equation}
where $M$ is a constant that represents the mass of the interior object, and $\zeta_e$ is a constant introduced for later convenience. This constant $\zeta_e$ guarantees that the time coordinate of both the interior wormhole and the exterior vacuum solutions is the same upon matching.

\subsection{Matching the wormhole interior to the vaccum exterior}

Let us now apply the junction conditions derived in Sec.\ref{sec:junct} to match the interior wormhole solution described by the line element in Eq.\eqref{metric} with redshift and shape functions given by Eqs.\eqref{redshift} and \eqref{shape}, respectively, to the exterior vacuum solution described by the line element in Eq.\eqref{schwarz}. 

The first junction condition, i.e., Eq.\eqref{junction1}, is used to set a value for the constant $\zeta_e$. As this condition imposes the continuity of the metric $g_{ab}$ across the hypersurface $\Sigma$, using Eqs.\eqref{metric} and \eqref{schwarz} we obtain a relationship between $\zeta_e$ and $\zeta_0$ as
\begin{equation}\label{setzetae}
e^{\zeta_0\left(\frac{r_0}{r_\Sigma}\right)^p}=\left(1-\frac{2M}{r_\Sigma}\right)e^{\zeta_e},
\end{equation}
where $r_\Sigma$ is the radius at which the matching between the two metrics is performed, which will be set by the second junction condition.

Let us now turn to the second junction condition, i.e., Eq.\eqref{junction2}. This junction condition consists of a constraint to the radius $r_\Sigma$ at which the matching must be done. For the two metrics in Eq.\eqref{metric} and \eqref{schwarz}, the junction condition $\left[K\right]=0$ takes the form
\begin{equation}\label{setrsigma}
\frac{2r_\Sigma-3M}{r_\Sigma^2\sqrt{1-\frac{2M}{r_\Sigma}}}=\frac{1}{2r_\Sigma}\sqrt{1-\left(\frac{r_0}{r_\Sigma}\right)^{q+1}}\left[4-p\zeta_0\left(\frac{r_0}{r_\Sigma}\right)^p\right].
\end{equation}
For each particular combination of parameters $p$, $q$, and $\zeta_0$, Eq.\eqref{setzetae} must be solved for $r_\Sigma$ and the matching between the two metrics must be performed at $r=r_\Sigma$. One can now insert the value of $r_\Sigma$ into Eq.\eqref{setzetae} and, for each combination of parameters, obtain the corresponding value of $\zeta_e$.

In this work, we are interested in isotropic perfect-fluid thin shells. In the perfect-fluid situation, the surface energy density $\sigma$ and the tangencial pressure $p$ of the thin-shell can be obtained from the stress-energy tensor $S_{\alpha}^\beta=S_{ac}e_\alpha^ae^c_\gamma h^{\beta\gamma}$ which takes the diagonal form
\begin{equation}\label{sabdiag}
S_\alpha^\beta=\text{diag}\left(\sigma,p_s,p_s\right).
\end{equation}
Using Eq.\eqref{stressts}, the surface energy density $\sigma$ and the transverse pressure $p_s$ of the thin-shell are thus given by
\begin{eqnarray}
&\sigma = \frac{\epsilon}{8\pi}\left[\left[K_0^0\right]\left(1+\gamma+2\alpha R_\Sigma\right)+\right.\nonumber\\
&\left.+2\alpha\left(K^{0(\Sigma)}_{0}\left[R\right]-n^c\left[\nabla_c R\right]\right)\right],\label{sigma}
\end{eqnarray}
\begin{eqnarray}
&p_s=\frac{\epsilon}{8\pi}\left[\frac{1}{2}\left[K_0^0\right]\left(1+\gamma+2\alpha R_\Sigma\right)+\right.\nonumber\\
&\left.+\alpha\left(K^{0(\Sigma)}_{0}\left[R\right]+2n^c\left[\nabla_c R\right]\right)\right],\label{press}
\end{eqnarray}
where we have used the fact that, since $\left[K\right]=0$ holds, then in spherically symmetric spacetimes and spherical coordinates we have $\left[K_0^0\right]=-2\left[K_\theta^\theta\right]=-2\left[K_\phi^\phi\right]$. Finally, using Eqs.\eqref{sigma} and \eqref{press}, one computes the relationship $\sigma+p$. For the NEC to be satisfied at the thin-shell, it is required that $\sigma+p>0$. We have now all the necessary tools to perform the matching and obtain the full wormhole solution. 

As an example, let us take the solution previously obtained in Sec.\ref{sec:wormsol} and plotted in Fig.\ref{fig:nec}. Using Eq.\eqref{setrsigma} with $p=1$, $q=1$, and $\zeta_0=-60$ we obtain $r_\Sigma=2.02346M$. Then, inserting this result into Eq.\eqref{setzetae}, we obtain $\zeta_e=-54.847$. Finally, using Eqs.\eqref{sigma} and \eqref{press} we obtain $8\pi(\sigma+p)=0.996129/M$. This solution features combinations $\rho+p_r$ and $\rho+p_t$ positive in the whole range of $r$ between $r_0$ and $r_\Sigma$, a vanishing $T_{ab}$ for $r>r_\Sigma$, and also $\sigma+p>0$ at the shell, and thus it consists of a solution that satisfies the NEC throughout the whole spacetime. Following the same methodology, a wide variety of other solutions satisfying these requirements can be obtained with different combinations of parameters. 

Regarding the extra terms on the stress-energy tensor arising from $\left[R\right]\neq 0$, i.e., $S_a$, $S$ and $s_{ab}$, since $\left[R\right]$ is a function of $r$ only due to the spherical symmetry of the problem, then the only non-zero component of $\nabla_c\left[R\right]$ is the radial component. As the induced metric projects tensors into a hypersurface of constant radius, the contraction $h_a^c\nabla_c\left[R\right]$ in Eq.\eqref{sa} vanishes and one obtains $S_a=0$. On the other hand, the scalar $S$ can be computed directly from Eq.\eqref{s}, from which one obtains $8\pi S=1.5145/M$.  The double-layer distribution $s_{ab}\left(l\right)$ is defined as in Eq.\eqref{sab} and requires some test function to be computed explicitly. In particular, since we are interested in spacetimes that satisfy the NEC, and thus we have to verify that $T_{ab}k^ak^b\geq 0$ for some null vector $k^a$, a particularly interesting test function $Y^{ab}$ to study in Eq.\eqref{sabfinal} is $Y^{ab}=k^ak^b$. Given that the spacetime is spherically symmetric and that $\Sigma$ is an hypersurface of constant radius, one can always find a coordinate transformation in such a way that the metric on $\Sigma$ reduces to a Minkowski spacetime. Considering thus a general null vector of the form $k^a=\left(-1,a,b/r, c/(r\sin\theta)\right)$, where the constants $a$, $b$, and $c$ satisfy the relationship $a^2+b^2+c^2=1$, one verifies that the factor $n^c\nabla_c\left(k^ak^b\right)$ vanishes for any $a$, $b$ and $c$, and thus the integral on the RHS of Eq.\eqref{sabfinal} vanishes and the double gravitational layer does not contribute to the NEC.

\section{Conclusions}\label{sec:concl}

In this work, we have derived asymptotically flat traversable wormhole solutions satisfying the NEC throughout the whole spacetime in a quadratic form of the generalized hybrid metric-Palatini gravity in the geometrical representation. The wormhole solutions consist of an interior wormhole solution with a non-exotic perfect fluid near the throat, an exterior Schwarzschild solution, and a double gravitational layer thin-shell at the separation hypersurface between the interior and exterior solution. 

We have shown that the general set of junction conditions previously derived in \cite{rosa1} can be simplified for particular forms of the function $f\left(R,\mathcal R\right)$. More precisely, if one selects a function $f$ that is quadratic in $R$ and linear in $\mathcal R$, the junction conditions $\left[R\right]=0$ and $\left[\mathcal R\right]$ cease to be mandatory. As a consequence, extra terms arise in the stress-energy tensor $S_{ab}$ of the thin-shell, thus giving rise to a double-layer thin-shell distribution at the junction hypersurface.

Unlike in the previously published general case for which the derivation of traversable wormhole spacetimes satisfying the NEC required fine-tuning \cite{rosa1}, in this situation the simplified set of junction conditions implies fewer restrictions on the solutions, and thus the method outlined in this work allows one to easily obtain numerous solutions and for a wide variety of parameters. A particularly interesting advantage is that asymptotic flatness can be preserved in this situation, unlike the general case where the validity of the NEC could only be guaranteed for asymptotically AdS spacetimes.

Although there is still no clear physical interpretation for the double-layer stress-energy tensor distribution, in this work we have explicitly computed the integral of this distribution over the entire spacetime for a particular test-function given by $Y^{ab}=k^ak^b$, where $k^a$ is a null vector. This choice of the test function was motivated by the NEC $T_{ab}k^ak^b\geq 0$, since the double-layer stress-energy tensor distribution appears naturally as an extra term in the stress-energy tensor $T_{ab}$. Given that the hypersurface $\Sigma$ is an hypersurface of constant radius, we have obtained that this integral vanishes and thus the double-layer stress-energy tensor does not contribute to the NEC. This result seems to apply to any static and spherically symmetric spacetime.

\begin{appendix}

\section{Quadratic GHMPG in the scalar-tensor representation}\label{sec:app}

The generalized hybrid metric-Palatini gravity, similarly to the $f\left(R\right)$ theories of gravity, can be rewritten in terms of a dinamically equivalent scalar-tensor theory via the definition of auxiliary scalar fields. This scalar-tensor representation of the theory was used e.g. in the context of cosmological solutions \cite{rosa3} and traversable wormholes \cite{rosa1}. Thus, the particular case of quadratic gravity should also be studied in this representation to emphasize the equivalence of the two formalisms. However, the particular form of the function $f$ associated with this example does not belong to the domain of functions for which the scalar-tensor representation is defined, as we show in the following.

Let us start by rewriting the action in Eq.\eqref{initialaction} with two auxiliary fields $\alpha$ and $\beta$, respectively, in the following form
\begin{eqnarray}
S=\frac{1}{2\kappa^2}\int_\Omega \sqrt{-g}\Big[f\left(\alpha,\beta\right)+\frac{\partial f}{\partial \alpha}\left(R-\alpha\right)
    \nonumber  \\
+\frac{\partial f}{\partial\beta}\left(\cal{R}-\beta\right)\Big]d^4x+S_m.\label{gensca}
\end{eqnarray}
This action is a function of three independent variables, namely the metric $g_{ab}$ and the two auxiliary fields $\alpha$ and $\beta$. Taking the variations of this action with respect to the fields $\alpha$ and $\beta$ yields the two equations
\begin{equation}
\frac{\partial^2 f}{\partial\alpha^2}\left(R-\alpha\right)+\frac{\partial^2 f}{\partial\alpha\partial\beta}\left(\cal{R}-\beta\right)=0,
\end{equation}
\begin{equation}
\frac{\partial^2 f}{\partial\beta\partial\alpha}\left(R-\alpha\right)+\frac{\partial^2 f}{\partial\beta^2}\left(\cal{R}-\beta\right)=0.
\end{equation}
These two coupled equations can be rewritten in a matrix form $\mathcal M \textbf{x}=0$ as
\begin{equation}
\label{matrix}
\mathcal{M}\textbf{x}=
\begin{bmatrix}
\frac{\partial^2 f}{\partial\alpha^2} & \frac{\partial^2 f}{\partial\alpha\partial\beta} \\[0.8em]
\frac{\partial^2 f}{\partial\beta\partial\alpha} & \frac{\partial^2 f}{\partial\beta^2} 
\end{bmatrix}
\begin{bmatrix}
R-\alpha \\[0.8em]
\cal{R}-\beta
\end{bmatrix}
=0
\end{equation}
The solution for Eq.~\eqref{matrix} is unique if and only if the determinant of $\mathcal M$ does not vanish, i.e., $\det \mathcal M\neq 0$. This condition yields the relation
\begin{equation}\label{ghmpgstequiv}
\frac{\partial^2 f}{\partial\alpha^2}\frac{\partial^2 f}{\partial\beta^2}\neq \left(\frac{\partial^2 f}{\partial\alpha\partial\beta}\right)^2.
\end{equation}
If the condition above is not satisfied, then the solution of Eq.~\eqref{matrix} is not unique and the scalar-tensor representation for such a function $f$ is not well-defined, and thus the equivalence between the two representations is not guaranteed. This happens because for these particular cases the definitions of the scalar fields as functions of $R$ and $\mathcal R$ are not invertible.

The form of the function $f$ for which we could drop the junction conditions $\left[R\right]=0$ and $\left[\mathcal R\right]=0$ is given in Eq.~\eqref{quadf}. For this function, we have $f_{\mathcal R\mathcal R}=0$ and $f_{\mathcal R R}=0$. This implies that the determinant of the matrix $\mathcal M$ vanishes, as
\begin{equation}
\text{det}
\begin{bmatrix}
f_{RR} & f_{R\mathcal R} \\[0.8em]
f_{R\mathcal R} & f_{\mathcal R\mathcal R}
\end{bmatrix}
=f_{RR}f_{\mathcal R\mathcal R}-f_{R\mathcal R}^2=0,
\end{equation}
and thus this form of the function $f$ does not have an equivalent counterpart in the scalar-tensor representation. Therefore, there are no particular cases of the scalar-tensor representation for which some of the junction conditions could be discarded.

\end{appendix}

\begin{acknowledgments}
We thank Gonzalo Olmo, Rui Andr\'{e} and Jos\'{e} P. S. Lemos for the discussion and suggestions. JLR was supported by the European Regional Development Fund and the programme Mobilitas Pluss (MOBJD647).
\end{acknowledgments}


\end{document}